# Physical Nature of Magnon Spin Seebeck Effect in Ferrimagnetic Insulators

Linjie Ding[a,b†], Dongchao Yang[a†], LiZhi Yi[a], Yunli Xu[a], Bingbing Zhang[a], Hua-Hua Fu[c], Shun-Qing Shen[d], Min Liu[a], Liqing Pan[a]*, and John Q. Xiao[e]*

a. Hubei Engineering Research Center of Weak Magnetic-field Detection, College of Science, China Three Gorges University, Yichang 443002, China
b. Department of Physics, Chongqing Three Gorges University, Chongqing 404100, China
c. School of Physics, Huazhong University of Science and Technology, Wuhan 430074, China
d. Department of Physics, University of Hong Kong, Hong Kong, China
e. Department of Physics and Astronomy, University of Delaware, Newark, DE, 19716, USA

The spin Seebeck effect (SSE) in ferrimagnetic insulators (FMI) provides a simple method of using heat to manipulate magnons, which could be used as carriers of information and energy conversion. However, a theory that can quantitively interpret experimental results is still lacking. In this paper, we develop a transport theory of magnons in FMI at low temperatures by combining the macroscopic Boltzmann equation with microscopic quantum scattering theory. It is found that the scattering of magnons is dominated by phonons rather than magnons, and the relaxation time of magnon is inversely proportional to the cube of temperature. At extremely low temperature region, the magnon enters the ballistic transport process. In addition, we also derive the linear spatial distribution of the transverse SSE signal with sample position. All the theoretical results are in excellent agreement with the experimental data.

*Corresponding author. Email: lpan@ctgu.edu.cn, jqx@udel.edu

†These authors contributed equally to this work.

## I. INTRODUCTION

Developing simple and powerful methods to control spin is a critical step to realize efficient spintronic devices [1]. Alternative to electrical means, using heat can also generate and manipulate spin current, which is known as the spin Seebeck effect (SSE) [2,3]. It refers to the phenomena of thermal-driven spin current from a ferromagnet into a neighboring nonmagnetic (NM) layer which converts the spin current into a detectable charge current due to the inverse spin Hall effect (ISHE). It's well known that spins can be transferred by the motion of electrons. But the first implement of SSE in a ferromagnetic insulator (FMI) [4], $Yi_3Fe_5O_{12}$ (YIG), confirmed that magnons, the quanta of spin waves, can also be used as spin transport carriers, and more importantly it is much more energy efficient to generate the magnon current than the electron spin current [5]. Up today, many ferro- or ferri- and antiferro-magnetic materials have been explored, like YIG, $NiFe_2O_4$ [6], and NiO [7], which are particularly interesting for energy efficient devices.

Experimentally, the SSE can be realized in two geometries according to the heat flow direction, the longitudinal (LSSE) and transverse (TSSE) configurations with temperature gradient perpendicular and parallel to the layers, respectively [8]. While, most experiments [9] have been performed using longitudinal configuration because of its simple experimental geometry and easy control of temperature gradient direction, satisfying theoretical interpretations of some observed experimental results in both configurations are still lacking.

The first unexplained result is the temperature-dependent LSSE signal in FMI/NM system [10,11]. The observation of LSSE first increases and then decreases with increasing temperature suggests multiple competing mechanisms, such as scatterings among magnons, phonons, grain boundaries and defects. Therefore, how to develop a theory to account for the temperature-dependent LSSE signal is still a challenge. Rezende *et al.*[12,13] first adopted Boltzmann transport approach to describe the temperature-dependent LSSE in YIG/Pt, but they only considered the contribution of magnon-magnon scattering, and the theoretical fitting deviates from the experimental data. Recently, magnon-phonon coupling has been directly observed through a resonant enhancement in the magnetic field dependence of LSSE [14], suggesting the importance of magnon-phonon interaction. Based on the previous work of Rezende *et al.*, Costa *et al.* [15] further introduced a damping factor $\alpha$ to include magnon dispersion through the magnon-phonon interaction, and numerically corrected the fitting curves to the experimental results. In addition, Jin *et al.* [16] derives linear temperature dependence of LSSE at low temperature region by considering the grain boundary scattering in FMI films, However, experimental data shows a $\sim T^{1.2}$ dependence.

The other unsolved problem is the spatial distribution of TSSE. The observed TSSE signal changes from positive to negative value almost linearly with the sample position from one end to the other with zero value in the middle of the sample [4]. It should be noted that TSSE experiments show poor repeatability from different research groups [17,18], and only a few groups can repeat the spatial dependent TSSE experiments [19,20]. This is probably due to the small TSSE signals (hundreds of nanovolts) or non-identical sample properties. Some literatures pointed out that the poor repeatability may be caused by the temperature difference between magnons and phonons in FMI [21]. But this explanation is inconsistent with a negligibly small temperature difference between magnons and phonons observed experimentally [22]. In addition, the temperature difference between magnons and phonons exists only within the magnon diffusion length near the boundary.

In this work, we focus on the transport properties of magnons in FMIs from the semi-classical Boltzmann equation to address these questions. And then, we reveal the microscopic scattering mechanisms of magnons. It was found that the temperature-dependent LSSE results at low temperature mainly depend on the magnon-phonon scattering and magnon-magnon scattering is secondary. The model of quantum transition probability establishes the relaxation time of magnon-phonon scattering follows $T^{-3}$ dependence, which precisely matches the experimental results. Furthermore, we came up with a new idea that the spatial distribution of TSSE can be theoretically explained by the diffusion of magnons.

The remainder of the paper is organized as follows. In Sec. II, we present the magnon transport theory at low temperature in FMIs first, and then reveal the scattering mechanism of magnon, after that, we do a comparison with the experiment results and account for the experimental observations. Finally, we summarize our conclusions in Sec. III.

## II. RESULTS AND DISCUSSION

### A. Magnon transport theory at low temperature in FMIs

First, we consider the thermal excitation of magnons. Assume there are $n_k$ magnons at the frequency of $\omega_k$, the average value of $n_k$ can be determined by the Planck distribution $\langle n_k \rangle = \left(e^{\hbar\omega_k/k_BT} - 1\right)^{-1}$. The total number of magnons can be written as $n = \int d\omega D(\omega) n(\omega)$, where $D(\omega)$ is the density of states for magnon. Using the dispersion relation at long wavelength approximation $\hbar\omega = Dk^2$ (see next section, or Ref.[13]), we obtain $n = C_1 (k_BT/D)^{3/2}$ and the average energy per magnon $\bar{U} = \left(C_2/4\pi^2 C_1\right)k_BT \approx 0.78 k_BT$, where $C_1 = \int_0^\infty dx \frac{x^{1/2}}{e^x - 1}$ and $C_2 = \int_0^\infty dx \frac{x^{3/2}}{e^x - 1}$. The average energy per magnon is $0.78 k_BT$. The actual average energy is lower because the part with energy higher than $k_B T_C$ should be removed.

A temperature gradient in FMI drives a magnon current which evolves from a non-equilibrium state to a steady state through scattering by magnons, phonons, lattice defects, and grain boundary. This steady state can be described as [23]

$$\frac{df}{dt} = \frac{\partial f}{\partial t} + \nabla_k f \cdot \frac{d\vec{k}}{dt} + \nabla f \cdot \frac{d\vec{r}}{dt} = \frac{\partial f}{\partial t}|_{coll} = -\frac{f_1}{\tau} \qquad (1)$$

with $f = f_0 + f_1$ where the equilibrium function $f_0$ obeys the Planck distribution, $f_1$ is the deviation from the equilibrium state. In a steady state, $\frac{\partial f_0}{\partial t} = 0$. Besides, there is no external force acts on magnons, thus $\hbar \frac{d\vec{k}}{dt} = \vec{F} = 0$. From the relaxation time approximation, the collision term can be written as $\frac{\partial f}{\partial t}|_{coll} = -\frac{f_1}{\tau}$. Under a temperature gradient, a steady state is reached through diffusion. We have $\nabla f_0 = \frac{\partial f}{\partial T} \nabla T$. As mentioned above, the magnon number $n$ depends on $T$, so the gradient of magnon number $\nabla n$ is produced by $\nabla T$. Thus, $\nabla f_0$ can also be written as $\nabla f_0 = \frac{\partial f}{\partial n} \nabla n_0 = \frac{\partial f}{\partial T} \frac{\partial T}{\partial n} \nabla n_0$, where $n_0$ is the magnon number at the steady state. The magnon current (in units of particle number) in FMI can be written as

$$J_m = \int -\tau_m \vec{v} \cdot \vec{v} (\frac{\partial f}{\partial T} \nabla T) d\vec{k} = -\left(\frac{C_3 k_B^{5/2}}{3\pi^2 \hbar^2 D^{1/2}}\right) \tau_m T^{3/2} \cdot \nabla T \equiv -D_m S_m \cdot \nabla T \qquad (2)$$

The magnon current $J_m$ is also represented by $J_m = -D_m \nabla n_0$. Here, $D_m = 2C_3 D \tau_m k_B T / 9\pi^2 C_1 \hbar^2$ is the diffusion coefficient of magnon and $S_m = 3C_1 k_B^{3/2} T^{1/2} / 2D^{3/2}$ is the SSE coefficient in FMI, where $C_3 = \int_0^\infty \frac{x^{5/2} e^x}{(e^x - 1)^2} dx$.

Compared with the SSE in ferromagnetic metals, the SSE coefficient for magnetic insulators can be denoted as $S_m$. If there are two kinds of magnons, spin-up and spin-down, then the definition of the SSE coefficient should be $S_S = \left( D_{m\uparrow} S_{m\uparrow} - D_{m\downarrow} S_{m\downarrow} \right) / \left( D_{m\uparrow} + D_{m\downarrow} \right)$. When the magnon of spin-down does not exist, then $D_{m\downarrow} = 0$, and $S_S = S_{m\uparrow} = S_m$.

The Equation (2) can be rewritten to include the magnon accumulation at the boundary or interface, $J_m = -D_m \nabla n = -D_m \nabla n_s - D_m S_m \nabla T = -D_m \nabla n_s - L_0 \nabla T$, where $n = n_0 + n_s$ with $n_s$ the magnon accumulation and $L_0 = C_2 \tau k_B^{5/2} T^{3/2} / 3\pi^2 \hbar^2 D^{1/2}$.

At a position far away from the boundary, the magnon current will not be affected by the magnon accumulation and is only determined by the applied temperature gradient. In this situation, $J_m = -D_m S_m \nabla T = -L_0 \nabla T$.

### B. Scattering mechanism of magnon in FMIs

From the previous section, we conclude that $J_m \propto \tau_m T^{3/2}$ through Boltzmann approach. The key parameter is magnon relaxation time $\tau_m$, which may come from four contributions including magnon-magnon ($\tau_{mm}$), magnon-phonon ($\tau_{mp}$), defects ($\tau_{md}$), and grain-boundary ($\tau_{mb}$) scatterings. The defect scattering has the same form as that from grain boundaries, and it can be included into $\tau_{mb}$. Therefore, the magnon relaxation time $\tau_m$ is expressed as:

$$\frac{1}{\tau_m} = \frac{1}{\tau_{mm}} + \frac{1}{\tau_{mp}} + \frac{1}{\tau_{mb}} \qquad (3)$$

Let us consider the magnons scattered by phonons in a ferrimagnetic insulator system containing phonons and magnons, like YIG, at low temperatures. The system Hamiltonian for a FMI is:

$$H = H_{magnon} + H_{ph} + H_{\text{int}} \qquad (4)$$

where the magnon Hamiltonian is $H_{magnon} = -2J_0 \sum_{\vec{l},\vec{\delta}} \vec{S}_{A,\vec{l}} \cdot \vec{S}_{B,\vec{l}+\vec{\delta}}$. After second quantization, it becomes

$$H_{magnon} = A + \sum_{\vec{k}} \left\{ \hbar\omega_{\vec{k}}^{(-)} \left( \alpha_{\vec{k}}^{\dagger}\alpha_{\vec{k}} + \frac{1}{2} \right) + \hbar\omega_{\vec{k}}^{(+)} \left( \beta_{\vec{k}}^{\dagger}\beta_{\vec{k}} + \frac{1}{2} \right) \right\} \tag{5}$$

where $\hbar\omega_{\vec{k}}^{(+)} = -4ZJ_0(S_a - S_b) - 8J_0 a^2 \frac{S_a S_b}{S_a - S_b} k^2$ for optical branch, while $\hbar\omega_{\vec{k}}^{(-)} = -8J_0 a^2 \frac{S_a S_b}{S_a - S_b} k^2$ for acoustical branch. At low temperatures, only the acoustical mode magnon is considered, i.e., $E(\vec{k}) = \hbar\omega_{\vec{k}}^{(-)} \sim Dk^2$.

The phonon Hamiltonian and the magnon-phonon interaction Hamitonian in Equation (4) are, respectively,

$$H_{ph} = \sum_{\vec{q}p} \hbar\omega_{\vec{q}p} \left( f_{\vec{q}p}^{\dagger} f_{\vec{q}p} + \frac{1}{2} \right) \tag{6}$$

$$H_{int} = -2J_0 \sum_{\vec{l},\vec{\delta}} \left[ \lambda \vec{e}_{\vec{l},\vec{l}+\vec{\delta}} \cdot \vec{u}_l \right] \vec{S}_{A,\vec{l}} \cdot \vec{S}_{B,\vec{l}+\vec{\delta}}, \tag{7}$$

where $\lambda$ is the magnon-phonon coupling strength. Here only single-phonon process is considered, and at the same time only the acoustical branch magnon is considered, i.e. α branch. For the acoustical branch of phonon, $\vec{u}_l = \frac{1}{\sqrt{N}} \sum_{\vec{q}p} \vec{e}_{\vec{q}p} \cdot G_{\vec{q}p} \left( f_{\vec{q}p}^{\dagger} + f_{-\vec{q}p} \right) e^{-i\vec{q}\cdot\vec{R}_l}$ with $G_{\vec{q}p} = \sqrt{\hbar/2M\omega_{\vec{q}p}}$ and $\omega_{\vec{q}p} = c_p q$ representing one longitude wave mode or two transverse wave modes.

Herein, $H_{int}$ is considered as a perturbation term. Thus, our main task is to calculate the square of the modulus of the first-order perturbation matrix element $H_{ij} \times H_{ij}^* = |H_{ij}|^2$. After the Holstein-Primakoff transformation and subsequently Bogoliubov transformation for operator diagonalization, only the bilinear term of the generation / annihilation operator is retained here, while other terms lead to $|H_{ij}|^2 = 0$.

$$H_{\text{int}} \approx \frac{1}{\sqrt{N}} \sum_{\vec{\delta}\vec{k}\vec{q}p} 2J \left[ e^{-i(\vec{k}-\vec{q})\cdot\vec{\delta}} - e^{i\vec{k}\cdot\vec{\delta}} \right] G_{\vec{q}p} \cdot \alpha^{\dagger}_{\vec{k}-\vec{q}} \alpha_{\vec{k}} \left( f^{\dagger}_{\vec{q}p} + f_{-\vec{q}p} \right) \tag{8}$$

where $J = -2J_0 \lambda \sqrt{S_a S_b}/(S_a - S_b)$, and $\Phi_{\vec{k}\vec{q}p} = \frac{1}{\sqrt{N}} \sum_{\vec{\delta}} 2J \left[ e^{-i(\vec{k}-\vec{q})\cdot\vec{\delta}} - e^{i\vec{k}\cdot\vec{\delta}} \right] G_{\vec{q}p}$.

Applying the Fermi golden rule, the reciprocal of the relaxation time for magnon-phonon scattering is:

$$\tau^{-1}_{\vec{k}\vec{q}p} = \frac{2\pi}{\hbar} |H_{\text{int}}|^2_{if} \delta(E_i - E_f) \tag{9}$$

Again using long wave approximation at low temperature, i.e., $E_{\vec{k}} \ll k_B T$, the reciprocal of the relaxation time of magnon after summation of all $\vec{k}$ is obtained

$$\tau^{-1}_m = \frac{32 S_a S_b J^2 a^2 k_m q_D}{\pi N M c_P^2} \cdot T^3 \propto T^3 \tag{10}$$

The results show that $\tau_m$ is proportional to $T^3$ at low temperatures. Early works about the magnon-phonon interactions had been reported in the literatures [24–26].

The magnon-magnon scattering has been discussed in great detail in Ref.[12,13], which cause the decrease of $l_m$ due to the increase of magnon number. However, after careful analysis below, we find that the magnon-phonon scattering is much stronger than the magnon-magnon scattering, and it is a dominant factor affecting the relaxation time of magnon [27].

Let us first look at the basic characteristics of phonons and magnons. For magnon, the density $n = C_1 (k_B T/D)^{3/2} T^{3/2}$, the total energy $U = 1.78(4\pi^2 D^{3/2})^{-1}(k_B T)^{5/2}$. Using $k_B T = 25 meV$ at 300 K, $D = 4.0 meV \cdot nm^2$, we have $n = 10^{21} cm^{-3}$ at 300 K and the average energy per magnon is $\bar{U} = U/n = 0.78 k_B T$. For phonons, the density $n_p = 2.4(2\pi^2 v^3 \hbar)^{-1}(k_B T)^3$, the total energy $U = \pi^4 (30\pi^2 v^3 \hbar^3)^{-1}(k_B T)^4$. For 300 K, $n_p = 6 \times 10^{23} cm^{-3}$, the average energy per phonon is $\bar{U} = U/n = 2.7 k_B T$. In the temperature range of 10-300 K, the number of phonons is much higher than that of magnons. Since the number of phonons and the average energy of phonons are much higher than those of magnons, the scattering of magnons in FMI will mainly come from magnon-phonon scattering.

It is note that the dependence of the number of magnons with temperature is $T^{-3/2}$. It can be roughly estimated that the relaxation time $\tau_{mm}$ due to the scattering of magnon follows the same power law with temperature. This means that the scattering between magnons would become remarkable at extremely low temperatures. However, at this time, the mean free path $l_m$ of magnon, due to the magnon-magnon scattering, is larger than the spacing between defects in the FMI thin films. The scattering effect between magnons will be submerged due to the scattering of defects and boundaries. The defect scattering (including grain-boundary scattering) dominates in this temperature region and the magnon's mean free path $l_m$ would be a constant $l_m = L$. Now magnon enters in the ballistic scattering region. Since $\tau = L/\bar{v}_m$, where $\bar{v}_m$ is the average velocity of magnon which shows $T^{1/2}$ dependence, $\tau_m$ has $T^{-1/2}$ dependence and $J_m \propto \tau_m T^{3/2} \propto T$.

### C. Comparison with experimental results in FMI/NM

#### 1. Theoretical calculation of LSSE in FMI/NM

In this section, we derive LSSE voltage theoretically in FMI/NM bilayer. The magnon or spin current distribution should be first determined. As shown in Equation (11a), the magnon accumulation obeys the diffusion equation. Equation 11b shows that the magnon current $J_m^F$ can be expressed as the sum of two contributions, the temperature gradient ($\nabla T$) term and magnon accumulation ($\nabla \mu_s$) term. In NM layer, the spin current obeys two current model for spin-up and spin-down electrons [28] as shown in Equation (11c).

$$\begin{cases} \nabla^2 \mu_s^F = \dfrac{\mu_s^F}{\lambda_m^2} & \text{(11a)} \\[6pt] J_m^F = -L_0 \nabla T - D_m \nabla n_s & \text{(11b)} \\[6pt] J_s^N = -\dfrac{1}{e}\nabla(\sigma_\uparrow \mu_\uparrow - \sigma_\downarrow \mu_\downarrow) - (\sigma_\uparrow S_\uparrow - \sigma_\downarrow S_\downarrow)\nabla T & \text{(11c)} \end{cases}$$

where $\mu_S^F$ and $\lambda_m$ are magnon accumulation, magnon diffusion length in FMI, respectively, and $\sigma_{\uparrow(\downarrow)}$, $\mu_{\uparrow(\downarrow)}$, $S_{\uparrow(\downarrow)}$ are conductivity, chemical potential, and Seebeck coefficient of NM for two spin channels, respectively. By applying the boundary conditions, the spin current distribution can be determined. Assuming a transparent interface, the spin current and chemical potential are continuous at the FMI/NM interface. The spin current equals zero at the film boundary. Here the chemical potential of magnons in FMI is $\mu_s^F = n_s \mu_0 / n$, where $\mu_0$ is the chemical potential at equilibrium, and $n$ is the number of magnons per unit volume. The coefficient 2e means the angular momentum of one magnon can be transferred to two electrons at the interface. By solving the above equations, the thermal-driven spin current distribution in FMI/NM can be obtained

$$2eJ_m^F = -2eL_0 \nabla T \left[ 1 - \dfrac{\cosh\left(\dfrac{z}{\lambda_F}\right)\sinh\left(\dfrac{d_N}{\lambda_N}\right)}{\dfrac{\sigma_{FMI}}{\sigma_N}\dfrac{\lambda_N}{\lambda_F}\cosh\left(\dfrac{d_N}{\lambda_N}\right)\sinh\left(\dfrac{d_F}{\lambda_F}\right) + \cosh\left(\dfrac{d_F}{\lambda_F}\right)\sinh\left(\dfrac{d_N}{\lambda_N}\right)} - \dfrac{\dfrac{\sigma_{FMI}}{\sigma_N}\dfrac{\lambda_N}{\lambda_F}\cosh\left(\dfrac{d_N}{\lambda_N}\right)\cosh\left(\dfrac{d_F+2z}{2\lambda_F}\right)\text{sech}\left(\dfrac{d_F}{2\lambda_F}\right)}{\dfrac{\sigma_{FMI}}{\sigma_N}\dfrac{\lambda_N}{\lambda_F}\cosh\left(\dfrac{d_N}{\lambda_N}\right) + \coth\left(\dfrac{d_F}{\lambda_F}\right)\sinh\left(\dfrac{d_N}{\lambda_N}\right)} \right],$$

(12a)

$$J_s^N = -2eL_0 \nabla T \dfrac{\sinh\left(\dfrac{d_N - z}{\lambda_N}\right)\tanh\left(\dfrac{d_F}{2\lambda_F}\right)}{\dfrac{\sigma_{FMI}}{\sigma_N}\dfrac{\lambda_N}{\lambda_F}\cosh\left(\dfrac{d_N}{\lambda_N}\right) + \coth\left(\dfrac{d_F}{\lambda_F}\right)\sinh\left(\dfrac{d_N}{\lambda_N}\right)},$$

(12b)

where $\sigma_{FMI} = 4nD_m e / \mu_0$ and $\sigma_N$ have the same dimension. For example, using the typical values of YIG at 300K, $\mu_0 = 0.77 k_B T$, $D = 5.32 meV \cdot nm^2$ [29], $\tau_m = 6 \times 10^{-11} s$ (using

$\tau_m = \lambda / v$, the spin diffusion length $\lambda_m = 1 \mu m$ [11] and $v = 1.6 \times 10^4 \, m/s$ taken from the parabolic dispersion relation), $\sigma_{FMI}$ of YIG can be obtained to be $7 \times 10^8 \, S/m$.

The electric field produced in the NM layer due to the ISHE can thus be obtained:

$$E_{ISHE} = \frac{\theta_{SH}}{\sigma_N} \frac{1}{d_N} \int_0^{d_N} J_s(z) dz = \frac{\theta_{SH}}{\sigma_N} \frac{\lambda_N}{d_N} \tanh(\frac{d_N}{2\lambda_N}) J_s(0) \,. \tag{13}$$

**Table 1** The spin Hall angle ($\theta_{SH}$), spin diffusion length ($\lambda_N$), conductivity ($\sigma_N$) of different detection materials taken from Ref. [30,31] and the calculated LSSE interfacial spin current $J_s(0)$ and ISHE electric field $E_{ISHE}$ with our model.

| NM | $\theta_{SH}$ | $\lambda_N$ (nm) | $\sigma_N$ ($\times 10^6$ S/m) | $J_s(0)$ (a.u.) | $E_{ISHE}$ (a.u.) |
|---|---|---|---|---|---|
| **Pt** | 0.1 | 7 | 2.08 | 6.8×10⁻³ | 1.6 |
| **Ta** | -0.07 | 1.9 | 0.34 | 6.7×10⁻³ | -4.5 |
| **W** | -0.14 | 2.1 | 0.56 | 8.5×10⁻³ | -7.4 |
| **Cu** | 0.0032 | 500 | 15.9 | 1.8×10⁻⁵ | 1.8×10⁻⁵ |
| **IrO₂** | 0.04 | 3.8 | 0.37 | 3.8×10⁻³ | 1.8 |

We applied this model to YIG/NM structures, where NM is Pt, Ta, W, Cu, and IrO$_2$ [30,31]. The used parameters are listed in **Table 1** and the numerical simulation of spin current distribution is plotted in Fig. 1. According to our calculation, the injected spin current at the interface can also be affected by the conductivity of the detection layer, like at the interfaces of YIG/Pt and YIG/IrO$_2$. In the past, to improve the LSSE signal, many efforts focused on finding powerful magnon source FMI materials [9] or increasing the spin-orbital coupling strength of the NM layer [32]. Few studies investigated the conductivity mismatch effect of FMI and NM, which would affect the interfacial spin current. Therefore, the mismatch of FMI and NM must be taken into account when optimizing the LSSE structure.

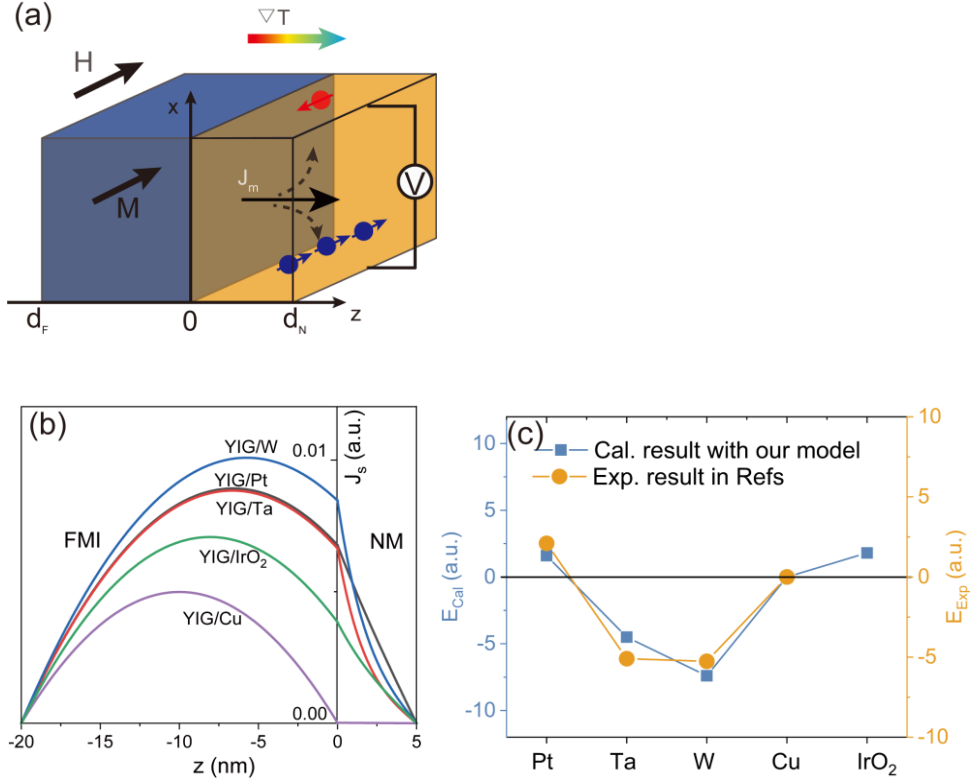

FIG. 1 (a) Schematic diagram of LSSE configuration. (b) shows the numerical simulation of spin current distribution in YIG(20)/NM(5nm) bilayers with different detection layers of Pt, Ta, W, Cu, and $IrO_2$, respectively. (c) Comparison between our calculated ISHE using Equation (13) and experimental data in Ref. [30].

## 2. Comparison with experimental LSSE results in FMI/NM

Overall, at low temperatures, we divide the magnon scattering into two regions: the ballistic scattering region at extremely low temperatures and the magnon-phonon scattering region otherwise, as shown in Fig. 2(a). In Fig. 2(b), the thermal-driven magnon current $j_m$ reaches a peak value with increasing temperature and then starts to decrease. The model depicted in section

B fits very well with the experimental data of YIG/Pt. We extracted the peak value for different YIG film thickness from ref. [11] and the results are shown in Fig. 2(c). We found the data obeys $T^{-3}$ law for thin films, which manifests the feature of temperature-dependent magnon relaxation time $\tau_{mp} \sim T^{-3}$, and the magnon mean free path $l_m$ is limited by the film thickness. When the film becomes too thick, the peak value of $l_m$ is determined by the defect spacing.

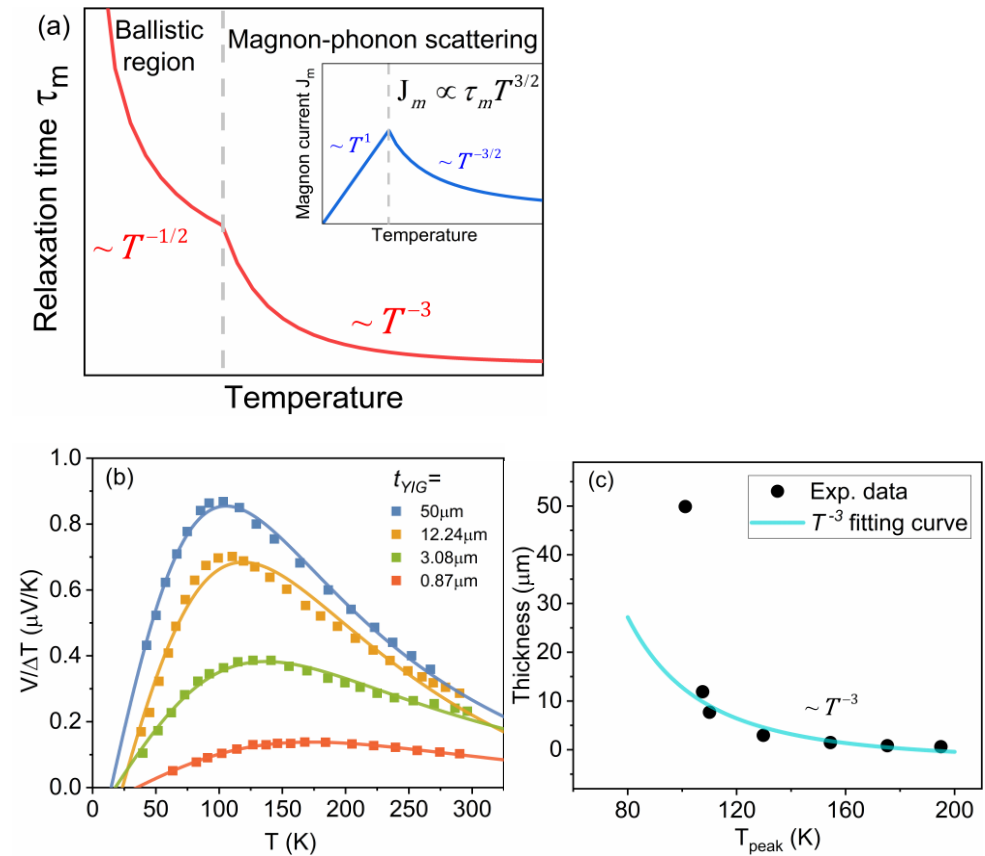

**Figure 2** (a) Theoretical model of temperature-dependent magnon relaxation time due to two main scattering events. Insert: the temperature-dependent magnon current. (b) Fitting curve calculated from the theory to the experimental data of temperature-dependent LSSE voltage. (c) The film thickness as a function of the temperature at peak position. The experimental data in (b) and (c) were taken from Ref. [11].

### 3. Explanation of Spatial-dependent TSSE

As shown in Fig. 3, the temperature gradient is along the *x* direction in TSSE configuration and the magnon current injected from FMI into NM is along the *z* direction. It is observed that the detected TSSE voltage changes sign as a function of the position. In this section, we interpret this phenomenon as the diffusion of magnons and give the expression of spatial-dependent magnon current in TSSE.

The magnon accumulation obeys the diffusion equation: $\nabla^2 \mu_s = \frac{\mu_s}{\lambda_m^2}$. Its general solution is $n_s = A\cosh(x/\lambda_m) + B\sinh(x/\lambda_m)$, where A and B are coefficients determined by the boundary conditions, i.e. $\vec{J}_m\big|_{x=\pm L/2} = 0$, where *L* is the length of FMI. The magnon distribution along *x* axis can be written as

$$n(x) = n_0 + n_s = n_0(0)\left[1 + \frac{3\nabla T}{2T_0}x - \frac{3\nabla T \lambda_m}{2T_0 \cosh(L/(2\lambda_m))}\sinh(x/\lambda_m)\right] \qquad (14)$$

where $n_0(0)$ and $T_0$ are magnon number and temperature at *x*=0 in the steady state.

Due to the characteristics of the magnon current, the diffusion coefficient and the spin Seebeck coefficient of FMI strongly depend on temperature, the divergence of the magnon current in the direction of the temperature gradient is not zero. However, in the steady state, there will be no magnon accumulation inside the bulk of FMI. Hence, a magnon current will be generated at the interface with NM along the *z*-direction.

The magnon current in the *x* direction is shown in Equation (2). If an NM is in contact with FMI in the *z*-direction, shown in Fig. 3(a), The flux of the magnon current of the rectangular micro element is zero due to the conservation of angular momentum，which is the unique feature of magnons different from phonons. So

$$J_{m,in}^{z} = -\nabla \cdot J_{m}^{x}(x) \cdot d = \left(1 - \frac{5}{2}\frac{\nabla T}{T_0}x\right)J^0 \qquad (15)$$

where $J^0 \equiv 3L_0 \cdot d \cdot \nabla T^2$. The magnon current generated in z direction varies linearly with x, depicted as $J_{m,in}^{z}$ shown in Fig. 3(a). If the NM thickness is extremely small, the magnon currents will all be reflected back. However, since the magnon energy is extremely small, the energy of the scatters (such as phonons or boundary) is significantly higher. This reflection becomes diffuse with random directions. Therefore, the reflected magnon current no longer has a linear distribution with x, but a uniform value, shown as $J_{m,ref}^{z} = J^0$ in Fig. 4(a). The net magnon flow $J_m^z$ in the z direction is equal to $J_{m,in}^{z} - J_{m,ref}^{z}$, shown as $J_m^z$ in Fig. 3(a). This result is consistent with the changing trend of the ISHE signal with x observed in the experiment [4]. A numerical simulation of the special distribution of magnon number in FMI as described by Equation (14) is plotted in Fig. 3(b). Even if the magnon mean free path $\lambda_m$ reaches up to $100\,\mu m$, the influence of the magnon accumulation at two ends of FMI is negligible, which results in an almost linear dependence of magnon number on position. Actually, the experimental value (measured by Brillouin light scattering, the original data adapted from Ref. [22] ), fitted with Equation (14) of magnon mean free path is only a few micrometers, as shown in Fig. 3(c).

The spatial-dependent TSSE experiments are not easy to repeat. To improve the experiment, we emphasize the attention must be paid to the following issues: (1) magnetization direction should be parallel to the temperature gradient along the *x*-direction; (2) the crystal quality of FMI should be excellent, so that the relaxation time of the magnon and the density of the generated spin wave are large which lead to larger ISHE; and (3) the interface and the quality of NM epitaxial film must be good and transparent to spin wave propagation.

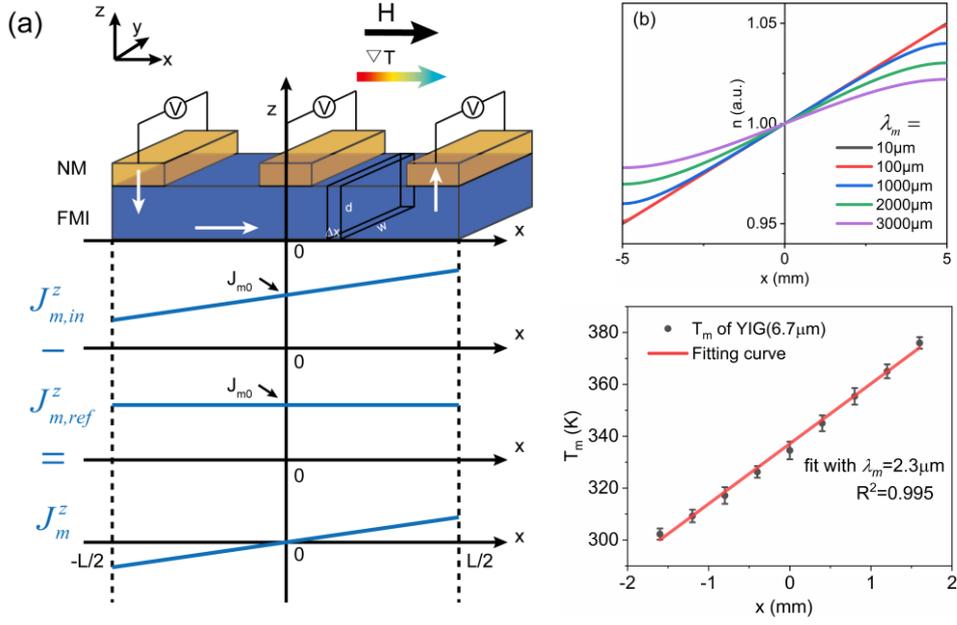

FIG. 3 (a) Schematic diagram of TSSE configuration and the calculated special variation of magnon current along *x* and *z* direction. (b) Calculated special-dependent distribution of magnon number with different mean free path. (c) Direct measurement of magnon temperature $T_m$, taken from Ref. [22].

### III. CONCLUSIONS

After we reviewed the recent research progress in magnon spin Seebeck effect in ferrimagnetic insulators, we have derived the transport equation of the magnon current driven by temperature gradient. After comparing and analyzing the results of previous important LSSE experiments, the magnon transport mechanism is fully revealed, *i.e.*, the relaxation time of the magnon is mainly determined by the magnon-phonon scattering and the magnon-grain-boundary scattering, rather than the scattering between magnons, and the scattering mechanism of $\tau_{mp} \sim T^{-3}$ is confirmed. In addition, we have also provided a theory that explains the position dependence of TSSE. The transverse distribution of the magnon temperature corresponding to the number of magnons is consistent with the experimental measurement, and due to the conversation of angular momentum of magnons, the linear spatial distribution of the transverse SSE signal with position is given. Our results provide new insights into understanding the magnon SSE experiments.

**Acknowledgments**

This work was supported by the National Natural Science Foundation of China (Grant No.: 51371105), J.Q.X. was supported by the National Science Foundation of the USA (Grant No.: DMR1505592). The authors particularly thank Prof. B. Hillebrands for supporting the original Brillouin light scattering data in Fig. 3(c).

**REFERENCES**

[1] R. Ramaswamy, J. M. Lee, K. Cai, and H. Yang, Appl. Phys. Rev., **5**, 031107 (2018).

[2] K. Uchida, S. Takahashi, K. Harii, J. Ieda, W. Koshibae, K. Ando, S. Maekawa, and E. Saitoh, Nature **455**, 778 (2008).

[3] K. Vandaele, S. J. Watzman, B. Flebus, A. Prakash, Y. Zheng, S. R. Boona, and J. P. Heremans, Mater. Today Phys. **1**, 39 (2017).

[4] K. Uchida, J. Xiao, H. Adachi, J. Ohe, S. Takahashi, J. Ieda, T. Ota, Y. Kajiwara, H. Umezawa, H. Kawai, G. E. W. Bauer, S. Maekawa, and E. Saitoh, Nat. Mater. **9**, 894 (2010).

[5] A. V. Chumak, V. I. Vasyuchka, A. A. Serga, and B. Hillebrands, Nat. Phys. **11**, 453 (2015).

[6] D. Meier, T. Kuschel, L. Shen, A. Gupta, T. Kikkawa, K. Uchida, E. Saitoh, J. M.

Schmalhorst, and G. Reiss, Phys. Rev. B **87**, 054421 (2013).

[7]  J. Holanda, D. S. Maior, O. Alves Santos, L. H. Vilela-Leão, J. B. S. Mendes, A. Azevedo, R. L. Rodríguez-Suárez, and S. M. Rezende, Appl. Phys. Lett. **111**, 172405 (2017).

[8]  K. Uchida, H. Adachi, T. Ota, H. Nakayama, S. Maekawa, and E. Saitoh, Appl. Phys. Lett. **97**, 172505 (2010).

[9]  K. I. Uchida, H. Adachi, T. Kikkawa, A. Kirihara, M. Ishida, S. Yorozu, S. Maekawa, and E. Saitoh, Proc. IEEE, **104**, 1946 (2016).

[10] T. Kikkawa, K. Uchida, S. Daimon, Z. Qiu, Y. Shiomi, and E. Saitoh, Phys. Rev. B **92**, 064413 (2015).

[11] E. J. Guo, J. Cramer, A. Kehlberger, C. A. Ferguson, D. A. MacLaren, G. Jakob, and M. Kläui, Phys. Rev. X **6**, 031012 (2016).

[12] S. M. Rezende, R. L. Rodríguez-Suárez, R. O. Cunha, A. R. Rodrigues, F. L. A. Machado, G. A. Fonseca Guerra, J. C. Lopez Ortiz, and A. Azevedo, Phys. Rev. B **89**, 014416 (2014).

[13] S. M. Rezende, R. L. Rodríguez-Suárez, R. O. Cunha, J. C. López Ortiz, and A. Azevedo, J. Magn. Magn. Mater. **400**, 171 (2016).

[14] T. Kikkawa, K. Shen, B. Flebus, R. A. Duine, K. Uchida, Z. Qiu, G. E. W. Bauer, and E. Saitoh, Phys. Rev. Lett. **117**, 207203 (2016).

[15] S. S. Costa and L. C. Sampaio, J. Phys. Condens. Matter **31**, 275804 (2019).

[16] H. Jin, S. R. Boona, Z. Yang, R. C. Myers, and J. P. Heremans, Phys. Rev. B **92**, 054436 (2015).

[17] M. Schmid, S. Srichandan, D. Meier, T. Kuschel, J. M. Schmalhorst, M. Vogel, G. Reiss,

C. Strunk, and C. H. Back, Phys. Rev. Lett. **111**, 187201 (2013).

[18] D. Meier, D. Reinhardt, M. van Straaten, C. Klewe, M. Althammer, M. Schreier, S. T. B. Goennenwein, A. Gupta, M. Schmid, C. H. Back, J. M. Schmalhorst, T. Kuschel, and G. Reiss, Nat. Commun. **6**, 8211 (2015).

[19] G. Siegel, M. C. Prestgard, S. Teng, and A. Tiwari, Sci. Rep. **4**, 4429 (2015).

[20] C. M. Jaworski, J. Yang, S. Mack, D. D. Awschalom, R. C. Myers, and J. P. Heremans, Phys. Rev. Lett. **106**, 186601 (2011).

[21] M. Schreier, A. Kamra, M. Weiler, J. Xiao, G. E. W. Bauer, R. Gross, and S. T. B. Goennenwein, Phys. Rev. B **88**, 094410 (2013).

[22] M. Agrawal, V. I. Vasyuchka, A. A. Serga, A. D. Karenowska, G. A. Melkov, and B. Hillebrands, Phys. Rev. Lett. **111**, 107204 (2013).

[23] L. Yi, D. Yang, M. Liu, H. Fu, L. Ding, Y. Xu, B. Zhang, L. Pan, and J. Q. Xiao, Adv. Funct. Mater. **30**, 2004024 (2020).

[24] K. P. Sinha, and U. N. Upadhyaya, Phys. Rev. **127**, 432 (1962).

[25] C. M. Bhandari, and G. S. Verma, Phys. Rev. **152**, 731 (1966).

[26] L. J. Cornelissen, K. J. H. Peters, G. E. W. Bauer, R. A. Duine, and B. J. van Wees, Phys. Rev. B **94**, 014412 (2016).

[27] S. Streib, N. Vidal-Silva, K. Shen, and G. E. W. Bauer, Phys. Rev. B, **99**, 184442 (2019).

[28] G. E. W. Bauer, E. Saitoh, and B. J. van Wees, Nat. Mater. **11**, 391 (2012).

[29] A. J. Princep, R. A. Ewings, S. Ward, S. Tóth, C. Dubs, D. Prabhakaran, and A. T. Boothroyd, npj Quantum Mater. **2**, 63 (2017).

[30] H. L. Wang, C. H. Du, Y. Pu, R. Adur, P. C. Hammel, and F. Y. Yang, Phys. Rev. Lett.

**112**, 197201 (2013).

[31]  K. Fujiwara, Y. Fukuma, J. Matsuno, H. Idzuchi, Y. Niimi, Y. Otani, and H. Takagi, Nat. Commun. **4**, 2893 (2013).

[32]  J. Sinova, S. O. Valenzuela, J. Wunderlich, C. H. Back, and T. Jungwirth, Rev. Mod. Phys. **87**, 1213 (2015).